\documentclass[sigconf,nonacm,screen]{acmart}
\AtBeginDocument{%
  }

\renewcommand\footnotetextcopyrightpermission[1]{} 
\usepackage{appendix}

\usepackage{graphicx}
\usepackage{microtype}
\usepackage{xspace}
\usepackage{subcaption}
\usepackage{wrapfig}
\usepackage{algorithm}
\usepackage{amsthm}
\usepackage{verbatim}
\usepackage{mathtools}
\usepackage{soul}
\usepackage{enumitem}
\usepackage{bbm}
\usepackage{multirow}
\usepackage{algpseudocode}
\usepackage{amsmath}
\usepackage{xcolor}
\usepackage{soul}

\newcommand{\acro}{\text{ABE-VVS}\xspace}

\newcommand{\stream}{\text{PC-Stream}\xspace}

\begin{document}

\title{ABE-VVS: Attribute-Based Encrypted Volumetric Video Streaming}

\author{Mohammad Waquas Usmani}
\email{mohammadwaqu@umass.edu}
\affiliation{%
  \institution{University of Massachusetts Amherst}
  \state{Massachusetts}
  \country{USA}
}
\author{Susmit Shannigrahi}
\email{sshannigrahi@tntech.edu}
\affiliation{%
  \institution{Tennessee Technological University}
  \state{Tennessee}
  \country{USA}
}
\author{Michael Zink}
\email{zink@ecs.umass.edu}
\affiliation{%
  \institution{University of Massachusetts Amherst}
  \state{Massachusetts}
  \country{USA}
}

\renewcommand{\shortauthors}{Usmani et al.}

\begin{abstract}


This work introduces ABE-VVS, a framework that performs attribute-based selective coordinate encryption in point cloud–based volumetric video streaming, enabling lightweight yet effective digital rights management (DRM). Instead of encrypting entire point cloud frames, ABE-VVS selectively encrypts subsets of spatial coordinates (X, Y, Z) or their combinations, reducing encryption and decryption time by up to 40\% and 75\%, respectively, compared to full-frame encryption.

To our knowledge, this is the first work to provide an end-to-end evaluation of a DRM-enabled secure point cloud streaming system. We deploy a point cloud video streaming pipeline on the CloudLab testbed and evaluate three HTTP-based Attribute-Based Encryption (ABE) granularities: ABE-XYZ, which encrypts all spatial coordinates; ABE-XY, which encrypts only X and Y; and ABE-X, which encrypts only X. These schemes are compared against conventional HTTPS/TLS-based secure streaming and an unsecured HTTP-only baseline. Our results show that ABE-based streaming reduces server CPU usage by up to 80\% and cache CPU usage by up to 63\%, achieving performance comparable to HTTP-only streaming while maintaining similar cache hit rates. Furthermore, ABE-XYZ and ABE-XY incur lower client-side rebuffering than HTTPS, while ABE-X achieves rebuffering-free playback comparable to HTTP-only. Although ABE-VVS increases client-side CPU usage, the overhead remains modest and does not impact streaming quality, while providing additional benefits such as simplified key revocation, elimination of per-client encryption, and reduced server and cache load.



\end{abstract}



\maketitle

\section{Introduction} \label{sec:introduction}

Augmented and Virtual Reality (AR/VR) are rapidly emerging as major trends in next-generation multimedia, offering levels of immersion far beyond conventional formats.
Unlike traditional 2D or 360$^\circ$ videos~\cite{8845574,usmani2025securingimmersive360video}, which restrict the viewer to three degrees of freedom (3DoF) and permit only rotational movement from a fixed position, volumetric AR/VR content enables six degrees of freedom (6DoF)~\cite{jin2024capturedisplaysurveyvolumetric,10.1145/3394171.3413639,10.1145/3625468.3647617,10.1145/3458305.3459595}. This allows users both to view in any direction and to move naturally within a virtual environment.

A key data representation powering these immersive experiences is the point cloud~\cite{10.1145/3458305.3459595,10.1145/3625468.3647617,jin2024capturedisplaysurveyvolumetric}. Point clouds describe 3D scenes using collections of points, each storing geometric coordinates \((x, y, z)\) and appearance attributes such as color \((r, g, b)\). Complex environments are constructed by assembling multiple 3D objects into a single point cloud frame. When these frames are played in sequence, they form a volumetric video~\cite{10.1145/3625468.3647617,10.1145/3394171.3413639,jin2024capturedisplaysurveyvolumetric,10.1145/3458305.3459595} typically viewed through head-mounted displays (HMDs).

However, delivering such immersive content introduces several challenges. Point clouds contain substantially more data than traditional video formats, leading to high computational requirements and the demand for high streaming bitrates. These demands can increase end-to-end latency, which is particularly problematic for immersive applications that rely on very low delay to maintain responsiveness~\cite{10.1145/3625468.3647617}. Elevated latency contributes directly to motion-to-photon delay~\cite{10.1145/3386290.3396933}, a major factor in cybersickness and a significant barrier in immersive experience~\cite{10.3389/frvir.2020.582204}. In addition, securing volumetric streams—through digital rights management (DRM), fine-grained access control, and protected distribution—adds yet another layer of computational overhead, making it even more difficult to meet stringent latency requirements.

While considerable research has advanced volumetric content streaming on head-mounted displays—ranging from improving quality of experience (QoE) through viewport prediction~\cite{9327920,10.1145/3386290.3396933} and adaptive quality streaming~\cite{10.1145/3343031.3350917} to progress in point cloud compression and encoding~\cite{8784814,10.1145/3304109.3306225}—there has also been notable development of end-to-end volumetric video delivery systems designed for scalability and low latency~\cite{10.1145/3625468.3647617,10.1145/3458305.3459595,10.1145/3386290.3396933}. Several efforts integrate DASH~\cite{sodagar2011mpeg} into volumetric streaming pipelines~\cite{Hosseini_2018,10.1145/3343031.3350917,10935962,10.1145/3712676.3718339}, and others incorporate machine-learning–based super-resolution~\cite{278306} to improve rendered quality.

In contrast, secure distribution and access control for volumetric video have seen far less attention~\cite{10.1145/3394171.3413639}. Most existing security mechanisms target isolated point cloud objects~\cite{LI2024103896,6975179,ANNABY2024171680,inproceedingsJin,10816523} rather than continuous volumetric video streams, leaving a significant gap in scalable DRM solutions for immersive media.

In this paper, we present Attribute-Based Encrypted Volumetric Video Streaming (\acro), a framework that applies attribute-based selective point-cloud encryption to balance security with low computational overhead tailored specifically for volumetric streaming. \acro builds upon prior work~\cite{10.1145/3704413.3765298} that enables encrypting coordinates within point cloud frames at varying levels. By encrypting only the targeted coordinates, \acro degrades and distorts the visual quality that prevents meaningful viewing by unauthorized users. Built on Attribute-Based Encryption (ABE)~\cite{CPABE}, \acro secures the content itself—both in transit and at rest—eliminating the need for separate transport-layer protection, consistent with prior work~\cite{10.1145/3712676.3714450,usmani2025securingimmersive360video}.\textit{ To the best of our knowledge, \acro is the first complete, DRM-enabled secure volumetric point-cloud streaming system.}

Within the \acro framework, the origin encrypts each point cloud using attribute policies, enabling the content to be cached or distributed from any location. Upon receiving the encrypted content, the client requests a decryption key from the license server, which issues a key containing the necessary attributes that enable the client to perform decryption. Time-based attributes can also be used to enforce automatic key expiration that revokes access without the need to re-encrypt the content. 

We evaluate \acro’s potential within a secure volumetric point cloud streaming system by testing multiple encryption granularities and comparing performance against conventional HTTPS based secure streaming as well as unsecured HTTP-only streaming. Our results demonstrate that \acro reduces server CPU usage by up to 80\% and reduces cache CPU load by up to 63\%, achieving similar efficiencies with unsecured HTTP-only streaming while preserving similar cache hit rates and rebuffering free client side playback. 

 We also observe that disk-based caching increases cache response time compared to a no-cache configuration, leading to higher rebuffering in some cases. This effect is mitigated by using a RAM-based cache, which significantly reduces response time and, consequently, rebuffering, while preserving results otherwise comparable to disk-based caching.
\section{Background and Related Work} \label{sec:background}

This section provides an overview of 3D point clouds, Attribute-Based Encryption, and recent work on point cloud encryption and point cloud streaming.

\subsection{3D Point Clouds}
\label{subsec:pointcloud-def}

Point clouds are composed of discrete 3D points, each represented by spatial coordinates \((x, y, z)\) and, in many cases, additional attributes such as color \((r, g, b)\) or surface normals \((n_x, n_y, n_z)\). These points collectively describe the geometry of an object or scene and are commonly stored as individual frames in the standard polygon file format (\texttt{.PLY})~\cite{ply}. Unlike conventional image or video data, point clouds provide a flexible, spatially detailed representation. When such frames are captured sequentially, they form a volumetric video, enabling full six degrees of freedom (6DoF) interaction for the user~\cite{10.1145/3394171.3413639}, compared to the three degrees of freedom available in 360-degree video experiences~\cite{usmani2025securingimmersive360video}.

\subsection{Point Cloud Security}
\label{subsec:pointcloud-related}

A variety of studies have explored security mechanisms for point clouds. Several approaches, such as~\cite{inproceedingsJin,LI2024103896,ANNABY2024171680}, rely on chaotic-map–based transformations, while others, including~\cite{6975179}, use permutation and geometric rotation for encryption. More recent methods~\cite{ANNABY2024171680,6975179,LI2024103896} emphasize on preserving dimensional or spatial consistency—ensuring that an object's scale, proportions, and overall structure remain intact—so that encrypted point clouds can coexist with unencrypted elements inside virtual environments. Despite these advances, most existing techniques target full object-level protection for standalone point clouds, offering strong cryptographic security but often incurring high computational cost.

For instance, decrypting a 100k-point cloud requires 8–9 seconds in~\cite{6975179}, 4–5 seconds in~\cite{ANNABY2024171680}, and roughly 30 ms for 40k points in~\cite{inproceedingsJin} (which scales to approximately 75 ms for 100k points). Since decryption occurs on the user side—often on resource-constrained VR and AR  headsets, these costs directly impact playback latency. In comparison, our method decrypts a 108k-point cloud in only 45 ms, even when all X, Y, and Z coordinates are processed. This represents a substantial computational improvement and makes our approach suitable for the low-latency requirements of volumetric video streaming, where 15–30 point cloud frames must be processed each second.

Unlike~\cite{10816523}, which uses ABE and watermarking to manage access and trace misuse in isolated 3D models—without examining the time cost of encryption or decryption—our work focuses on volumetric video, providing low-latency encryption and practical DRM across continuous point cloud frames.

Another work, VVSec~\cite{10.1145/3394171.3413639} applies irreversible adversarial perturbations to distort volumetric video, incurring 3–9\,s per frame, whereas our approach enables efficient, fully reversible restoration for authorized users.

\subsection{Point Cloud Video Streaming}
\label{subseec:relatedworks-streaming}
Hosseini and Timmerer introduced DASH-PC~\cite{Hosseini_2018}, the first system to adapt MPEG-DASH for dynamic point cloud streaming of a single point-cloud object, enabling adaptive, per-frame streaming. They generate multiple Level-of-Density (LoD) representations through density sub-sampling and describe these in an MPD-like manifest, allowing clients to switch quality levels based on bandwidth, similar to traditional ABR streaming.

Van der Hooft et al. introduced PCC-DASH~\cite{10.1145/3343031.3350917}, highlighting and addressing the limitations of DASH-PC’s frame-based and single-object design. Their system uses segment-based streaming for multi-object volumetric scenes, with each object represented as a DASH AdaptationSet in the MPD and annotated with its position in the 3D scene. They further employ MPEG V-PCC compression to create multiple bitrate/quality levels, and introduce additional rate-adaptation mechanisms based on the user’s 6DoF position and viewport, determining which objects to fetch and at what quality. They later conducted subjective and objective QoE evaluations of PCC-DASH in~\cite{9123081}.

More recently, Chujo et al.~\cite{10935962} proposed a perceptual-quality-driven DASH streaming approach for point clouds. Orthogonal to DASH-centric efforts, Zhang et al. introduced YuZu~\cite{278306}, which incorporates super-resolution into the volumetric video pipeline: the system streams downsampled versions to save bandwidth and performs client-side upsampling using an ML-based super-resolution model, effectively acting as an SR-aware volumetric ABR system with demonstrated QoE gains. Most recently, Heidarirad and Wang introduced VV-DASH~\cite{10.1145/3712676.3718339}, an end-to-end, codec-agnostic DASH framework for volumetric video. In addition, Liu et al.~\cite{10108421} proposed using client-side caching to buffer repeated tiles in volumetric video streams to eliminate redundant network transfers.

None of these systems considers securing the streaming pipeline or providing any form of DRM. By contrast, our work explicitly introduces encryption into the volumetric streaming pipeline. \acro performs frame-based streaming of point clouds, without any compression, viewport adaptation, or other performance optimizations, creating a worst-case streaming scenario. This allows us to isolate and quantify the impact of introducing a selective attribute-based encryption layer on end-to-end latency, cacheability, and QoE. The optimization techniques developed in prior works are orthogonal to our approach and can be integrated into our pipeline to further improve efficiency. To the best of our knowledge, we are the first to treat encryption as a first-class component of a volumetric streaming system.

\subsection{Distributed \& Secure Point Cloud Streaming}

While our streaming client currently uses a single-bitrate configuration and performs frame-based delivery, \acro is fully compatible with the volumetric DASH-ABR techniques described in Sect.~\ref{subseec:relatedworks-streaming}, and such optimizations can be readily integrated into our pipeline to further improve efficiency and adaptivity.

\textbf{HTTP Secure (HTTPS)} extends HTTP by encrypting communication using TLS~\cite{rfc8446}, where a short-term session key is negotiated through long-term public and private keys to secure data exchange. A key limitation of HTTPS is its host-centric, point-to-point design: content must be encrypted separately for each receiver, increasing computational load on servers and reducing the effectiveness of caching, as we will demonstrate in the evaluation of \acro.

\textbf{Content Distribution Networks (CDN)} improve large-scale media delivery by caching and serving content from geographically distributed edge nodes. This reduces latency, lowers origin-server load, and provides more stable performance during periods of high demand. While originally designed for conventional video, the same infrastructure can support volumetric point cloud video distribution: each point cloud frame can be cached and delivered as an independent object. Caching can be potentially improved by tailoring admission and eviction algorithms to the characteristics of point clouds, but such work is beyond the scope of this paper.

Caching for plain HTTP is simple because intermediaries can store and forward content directly. However, HTTPS complicates caching since CDN nodes must perform TLS/SSL termination~\cite{10.1007/978-3-031-28486-1_27}. When a client makes a request, the CDN negotiates a secure channel, checks if the requested content is cached, and—if it is missing or stale—retrieves it securely from the origin server. Before serving it to the client, the CDN must re-encrypt the content and create a fresh TLS session~\cite{10.1007/978-3-031-28486-1_27,usavps}. This involves decrypting and re-encrypting data at every hop, which adds noticeable computational overhead.

\subsection{Attribute-Based Encryption (ABE):}
\label{sec:abe-intro}
Attribute-Based Encryption (ABE) is a public-key cryptographic approach that supports fine-grained, scalable access control. Rather than relying on explicit user identities, ABE issues keys and enforces access rules based on descriptive attributes such as a user’s role, affiliation, or region. In ABE, each ciphertext includes an access policy $Q$, and users obtain private keys annotated with their attribute sets. Decryption is possible only when the user’s attributes satisfy the embedded policy. \textit{This design enables adaptable key management, allowing attributes and user keys to be modified or revoked without re-encrypting the underlying data.}


\subsubsection{\textbf{Access Control Policy:}} Consider a VR data-sharing platform that needs to regulate access to immersive 3D content.  

\textbf{(a) \emph{Example Policy:}} Access is granted only to users who have the \texttt{Researcher} role and who are either affiliated with \texttt{University~X} or located in \texttt{Asia}.  

\textbf{(b) \emph{User Attributes:}} Each user is issued a private key $SK$ containing their attribute set. For instance, \textbf{Alice} holds \{\texttt{\textbf{Role}: Researcher}, \texttt{\textbf{Affiliation}: UnivX}\}, whereas \textbf{Bob} has \{\texttt{\textbf{Role}: Student}, \texttt{\textbf{Region}: Asia}\}. 

\textbf{(c) \emph{Encryption and Decryption:}} The content provider encrypts the VR dataset under public key $PK$ with the policy: “\textbf{Role} = Researcher AND (\textbf{Affiliation} = UnivX OR \textbf{Region} = Europe).’’ Only users whose attributes satisfy this rule—such as Alice—are able to decrypt the dataset.




\subsubsection{\textbf{Key Revocation and Updates:}} One advantage of ABE is its ability to handle revocation and attribute changes without re-encrypting stored data. For example, if Alice leaves the research group, her key $SK$ can be revoked or assigned an expiration time. Time-based revocation is commonly implemented by adding an expiration attribute to user keys, as demonstrated in~\cite{reddick2022wip,reddick2021design,reddick2022case}. Administrators can periodically refresh or invalidate these attributes as required.

By embedding access policies directly into the encrypted content, ABE enables the efficient management of user onboarding, role transitions, and regional restrictions through key updates, rather than costly data re-encryption, making it a strong fit for distributed VR and 3D point cloud content platforms.

\subsection{ABE for Streaming:} 
\label{subsec:ABE-works}
Prior work~\cite{10.1007/978-3-319-32689-4_26,10.1145/3712676.3714450,usmani2025securingimmersive360video} has applied ABE to video streaming, focusing mainly on traditional 2D and 360$^\circ$ content. Studies such as~\cite{10.1145/3712676.3714450,usmani2025securingimmersive360video} demonstrate that using ABE over HTTP can lower cache CPU load while preserving QoE when compared to HTTPS. Although~\cite{usmani2025securingimmersive360video} also explores frame-level selective encryption, these techniques are designed for the structure of compressed video streams and do not translate to point cloud formats. 

Prior work~\cite{10.1145/3704413.3765298} introduced a novel attribute-based selective coordinate encryption technique for point cloud data, whose representation and encoding differ fundamentally from traditional video formats and their corresponding encryption methods. Unlike prior ABE-based approaches, this approach operates directly on the spatial geometry of point clouds by encrypting selected X, Y, Z coordinates, enabling a fine-grained and flexible encryption granularity. Building on this foundation,~\cite{usmani2025secureaidrivensuperresolutionrealtime} extends the approach with point cloud frame downsampling to reduce bandwidth and latency, and using machine-learning–based super-resolution to reconstruct high-quality content at the client.

\acro builds on prior work~\cite{10.1145/3704413.3765298} but differs by being the first to evaluate a novel end-to-end secure volumetric streaming pipeline in which encryption is a first-class component, including an analysis of ABE-enabled caching and its impact on computational load at intermediate caches.



\section{System Architecture} \label{section:system}
\begin{figure}[!ht]\centering
  \includegraphics[width=1\columnwidth]{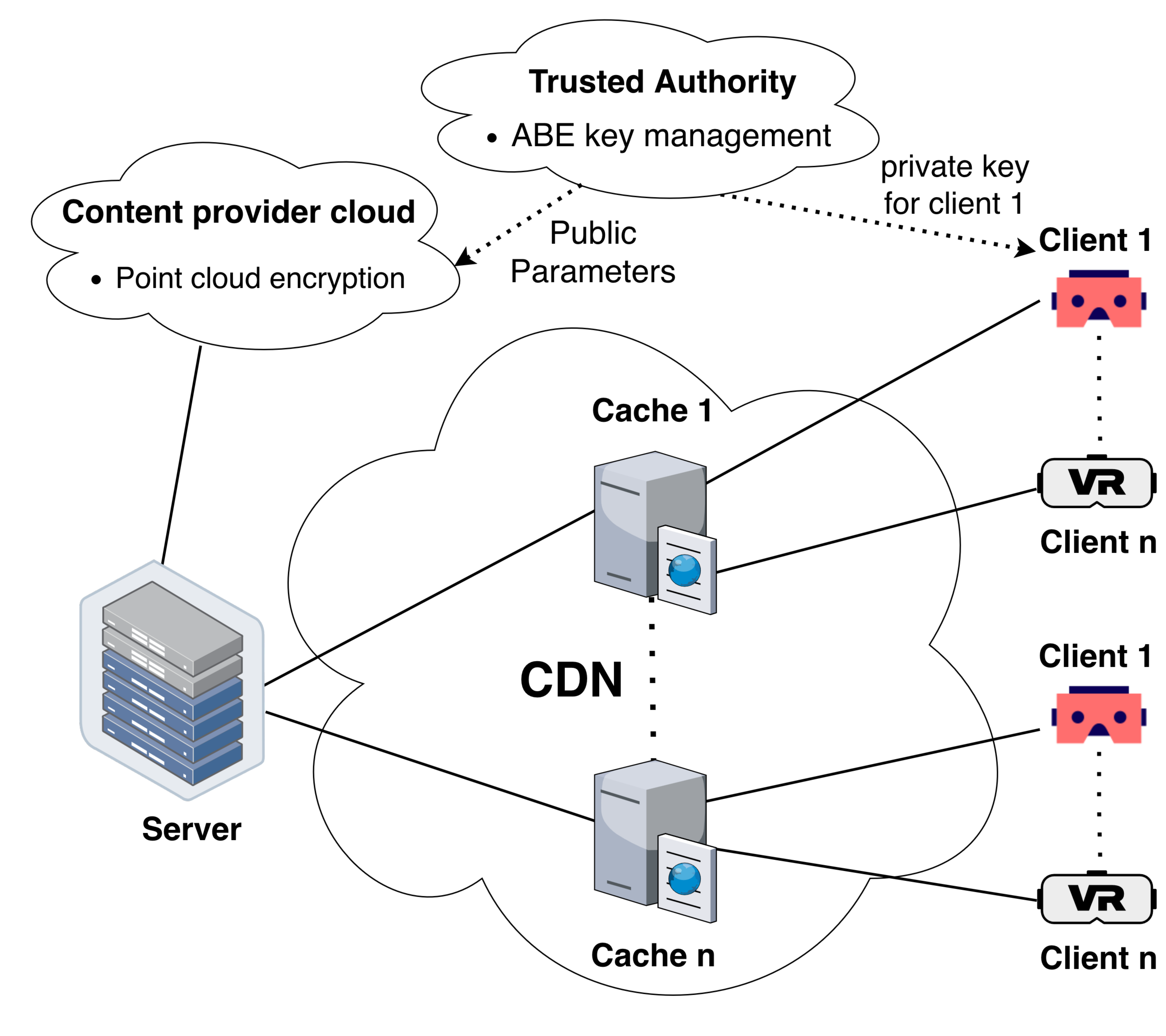} 
  \captionsetup{justification=centering}
  \caption{\acro System Architecture} 
  \label{fig:sys-arch} 
\end{figure}

In this section, we describe the overall system architecture of \acro and outline the key design choices behind its major components. Figure~\ref{fig:sys-arch} shows a simplified representation of such point cloud video distribution pipeline, highlighting the core elements relevant to our framework.

\subsection{HTTP-Based Point Cloud Streaming}
Most large-scale video platforms rely on CDNs and DASH/HLS-style ABR delivery. Our goal is to develop a volumetric point cloud streaming system that naturally fits within this ecosystem while supporting attribute-based selective coordinate encryption. To this end, we build a DASH-like player that parses an MPD, downloads frames via HTTP, HTTPS, or our HTTP-ABE mode, and maintains a playback buffer. Although our current prototype operates at a single quality level for evaluation purposes, the design is compatible with future ABR extensions.

Our approach build on a selective encryption mechanism tailored to volumetric content (Sect.~\ref{subsec:implement-enc}). Unlike conventional HTTP-based delivery, where content is either fully encrypted (HTTPS) or entirely unencrypted (HTTP), we encrypt only targeted coordinates within each point cloud frame. To signal these encrypted components, we extend the MPD with an ``\textit{encryption level}" attribute that specifies which coordinate subsets were protected. The client-side player reads this metadata and performs the correct selective ABE decryption scheme and reconstructs the point cloud frame.

A second requirement is secure key distribution. In contrast to HTTPS, where TLS implicitly handles key exchange, our system relies on a Trusted Authority (TA) that issues ABE keys to both the content provider and authorized clients. Caches remain oblivious: they store and serve encrypted point clouds without performing any cryptographic operations or participating in key management. We discuss the \acro prototype used in our evaluation in Sect.~\ref{subsec:prototype}.

\subsection{Selective Coordinate Encryption of Point Clouds}

\subsubsection{\textbf{Methodology:}}\label{subsec:implement-enc}In this work, we adapt the selective coordinate encryption methodology for point clouds proposed in~\cite{10.1145/3704413.3765298}. The primary objective is to reduce computational overhead by encrypting only a subset of point cloud coordinates rather than the entire data stream. \textit{Instead of aiming for complete privacy, the goal is to introduce sufficient visual distortion and obfuscation to prevent meaningful unauthorized viewing while maintaining efficiency.}


Each point cloud frame, stored in the \texttt{PLY} format, is parsed, and the selected coordinates are encrypted using the Ciphertext-Policy Attribute-Based Encryption (CP-ABE) toolkit~\cite{CPABE}. While the current implementation targets the \texttt{PLY} format, the same selective encryption mechanism can be readily extended to other point cloud representations.

\begin{figure*}[!ht]
\centering
\begin{minipage}[t]{1\columnwidth}
    \centering
    \includegraphics[width=\linewidth]{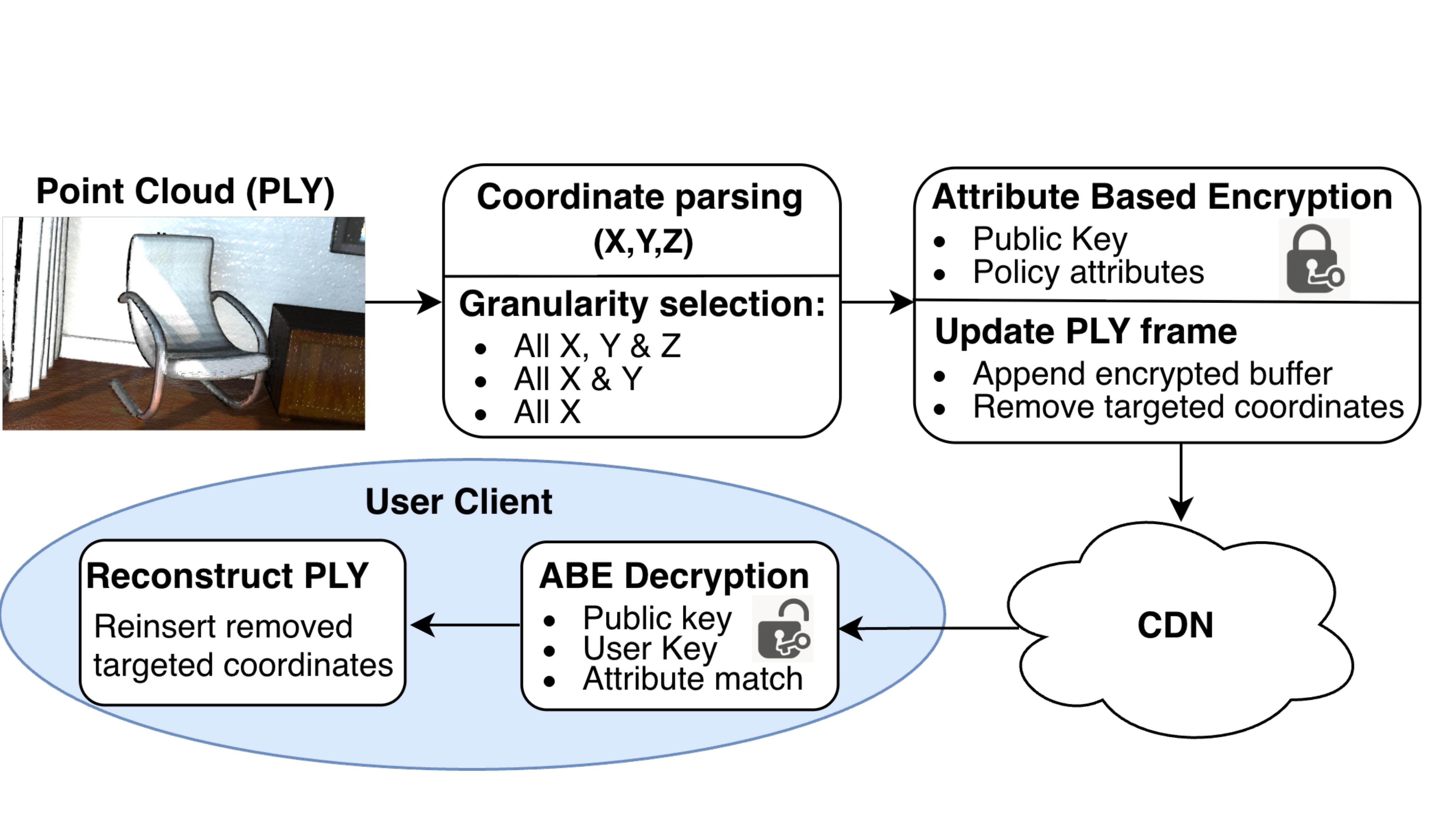}
    \captionsetup{justification=centering}
    \caption{Selective Coordinate Encryption Process}
    \label{fig:selective-encryption}
\end{minipage}\hfill
\begin{minipage}[t]{1\columnwidth}
    \centering
    \includegraphics[width=\linewidth]{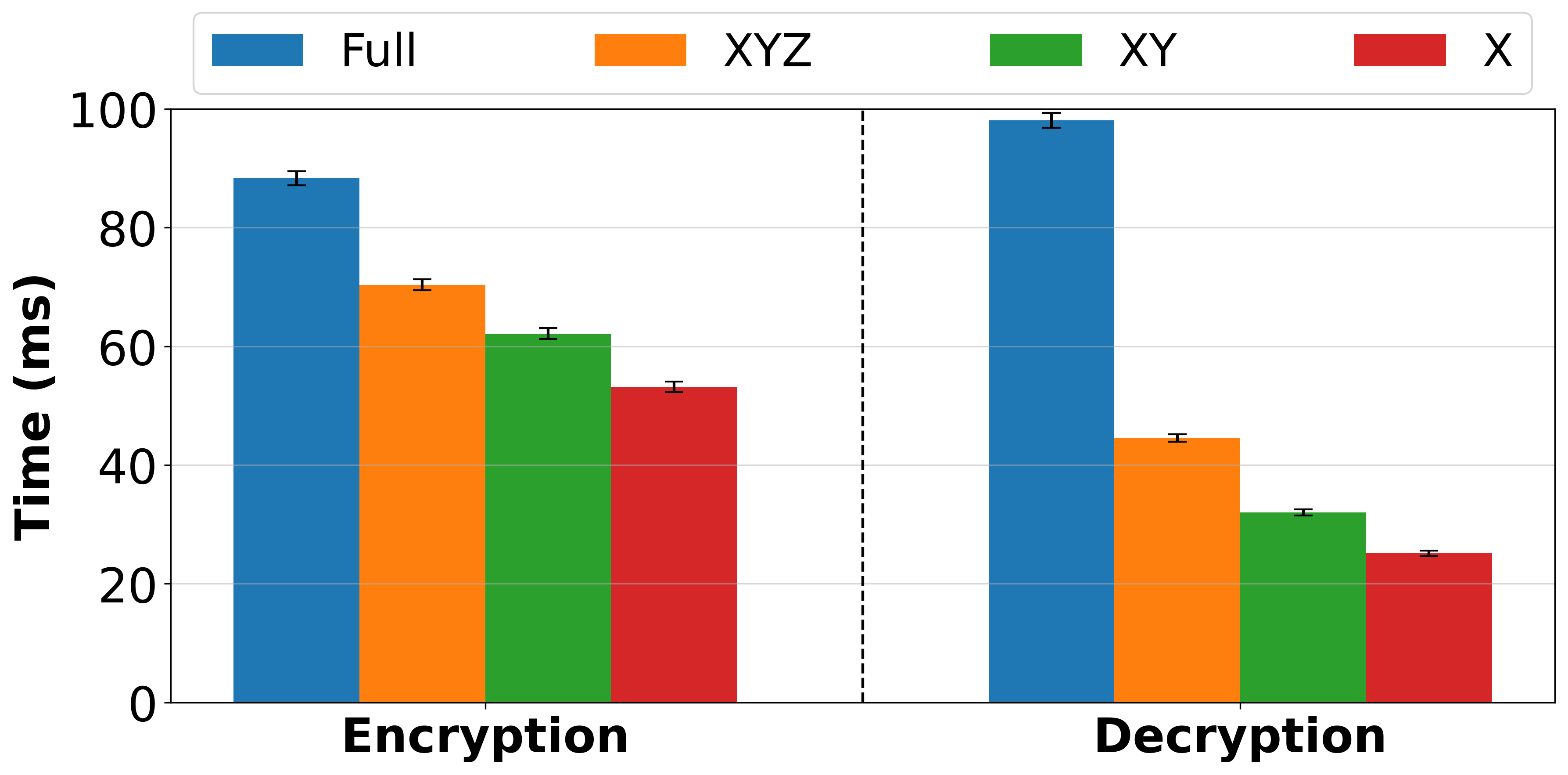}
    \captionsetup{justification=centering}
    \caption{Encryption and Decryption Times}
    \label{fig:enc-dec-times}
\end{minipage}
\end{figure*}

Figure~\ref{fig:selective-encryption} illustrates the overall workflow of our selective-coordinate encryption pipeline.
The encryption process begins by parsing the point cloud and extracting the coordinates of each vertex. A specified encryption granularity determines which coordinates are selected for encryption (e.g., all X, Y, Z values or only a subset), enabling flexible exploration of trade-offs among security, bandwidth overhead, and visual quality. The selected coordinates from each vertex are aggregated into a buffer and encrypted with ABE using the public key and access policy. The resulting ciphertext is appended to the point cloud frame, while all remaining data—such as headers, surface normals, and color attributes—remain unencrypted.



\textit{Unlike the approach in~\cite{10.1145/3704413.3765298}, which replaced encrypted coordinates with zero values for evaluation purposes, our implementation removes the targeted coordinates entirely from the vertex data. This modification avoids unnecessary bandwidth overhead, better suited for efficient streaming scenarios.}


Upon receiving the encrypted frame over the network, the client locates the encrypted buffer and decrypts it only if authorized, i.e., when its user key satisfies the attributes specified in the access policy. The client then reinserts the recovered coordinate values into their original positions within the vertex array, thereby fully reconstructing the original point cloud frame.

        

Encrypting a larger set of coordinates (e.g., all X, Y, Z) results in stronger visual obfuscation and higher security compared to encrypting fewer coordinates (e.g., only X). The work in~\cite{10.1145/3704413.3765298} provides a detailed evaluation of the visual distortion achieved under different selective encryption granularities; we refer interested readers to that study for an in-depth analysis of obfuscation characteristics.




\subsubsection{\textbf{Computational Analysis:}}\label{subsec:enc-impact} We evaluate the performance of the selective coordinate encryption approach with the targeted coordinates removed. All experiments are conducted on a CloudLab testbed node~\cite{duplyakin2019cloudlab} equipped with an \texttt{Intel Xeon E5-2630v3} processor, consisting of two sockets with 8 cores per socket and 2 hardware threads per core, and 64~GB of DRAM.
To ensure consistency, the same CloudLab cluster is subsequently used for the end-to-end streaming evaluation.


\textbf{Sample Point Cloud:} We use a point cloud containing 108,161 points (\textit{108k}) from the Open3D Office dataset~\cite{Open3D}, stored in the \texttt{PLY} format and acquired via RGB-D captures. Each point includes both geometric and appearance attributes: spatial coordinates \((x, y, z)\), surface normals \((n_x, n_y, n_z)\), and color values \((r, g, b)\). The coordinate and normal attributes are represented as double-precision floating-point values (8 bytes each), while color channels are stored as unsigned 1-byte integers.

\textbf{Encryption Granularities:} We evaluate the computational overhead of ABE encryption and decryption under different selective coordinate granularities. The evaluated configurations include: \texttt{\textbf{Full}}, which encrypts the entire frame without selection and serves as a baseline; \texttt{\textbf{XYZ}}, which encrypts all spatial coordinates; \texttt{\textbf{XY}}, which encrypts only the X and Y coordinates; and \texttt{\textbf{X}}, which encrypts only the X coordinate. These same granularities are later used in the streaming evaluation (Sect.~\ref{subsec:streaming}).

For clarity, we employ a single-attribute ABE policy in these experiments. While streaming deployments may require up to five attributes, prior work~\cite{reddick2022aabacautomatedattribute} demonstrates that ABE encryption time scales linearly with the number of attributes, and the increase from one to five attributes incurs negligible overhead.



\textbf{Encryption and Decryption times:}  We performed 1000 runs per encryption granularity and recorded the encryption and decryption times for the \textit{108k}-point cloud.

Figure~\ref{fig:enc-dec-times} reports the average runtimes in milliseconds. For encryption, relative to \texttt{Full} encryption, the \texttt{XYZ} granularity achieves up to a 20\% reduction in runtime, \texttt{XY} achieves up to 30\%, and encrypting only \texttt{X} yields approximately 40\%.

Decryption exhibits even larger improvements: we observe up to 55\% runtime reduction for \texttt{XYZ}, 67\% for \texttt{XY}, and 75\% when only \texttt{X} coordinates are decrypted. Notably, decryption benefits more from selective encryption than encryption itself. This is partly because full-frame decryption is inherently more expensive than full-frame encryption; however, across all selective granularities (\texttt{XYZ}, \texttt{XY}, and \texttt{X}), decryption consistently remains faster than encryption.

This asymmetry is advantageous in practice. Decryption occurs on client devices—often resource-constrained VR/AR head-mounted displays—where minimizing latency is critical, while encryption is performed by content providers with substantially greater computational resources. Finally, prior work~\cite{10.1145/3704413.3765298} has shown that both encryption and decryption times scale approximately linearly with point cloud size, suggesting that these benefits persist for larger volumetric content.

\subsection{Prototype Implementation}\label{subsec:prototype}
In this section, we present the implemented prototype of \acro using the following components:

\subsubsection{\textbf{Ciphertext-Policy Attribute-Based Encryption Toolkit:}}
We implement ABE using the CPABE toolkit~\cite{CPABE}, which follows the scheme introduced in~\cite{4223236}. The setup phase produces a master key ($MK$) and a public key ($PK$). The $PK$ is used to encrypt data under an access policy $Q$ that specifies the attribute conditions required for decryption. User keys ($SK$) are generated from the master key and the public key, and each $SK$ is associated with a set of attributes. Decryption succeeds only when the attributes encoded in the user's $SK$ satisfy the policy $Q$ embedded in the ciphertext.
The selective encryption process described in Sect.~\ref{subsec:implement-enc} is made possible by extending the CP-ABE toolkit to support selective coordinate encryption and decryption of point cloud frames.

\subsubsection{\textbf{PC-Stream client:}}\label{subsubsec:prototype:PC-Stream} We implemented a lightweight per-frame point cloud streaming client \textit{\stream} inspired by prior work~\cite{Hosseini_2018}. \textit{\stream} accepts an MPD file and parses it to obtain frame rate, per-frame point cloud URLs, and the encryption level indicating the coordinate granularity used during encryption. Both HTTP and HTTPS are supported, as identified by the MPD URL scheme.

The client maintains a configurable playback buffer, expressed in seconds. For example, a value of 2 permits up to two seconds worth of frames to be queued; if full, the client polls every 1ms for available space. This buffer also serves as the initial startup buffer before playback begins.

We integrate the attribute-based selective coordinate decryption module directly into the client. When the decryption option is enabled, the client decrypts the encrypted coordinate buffer by using the supplied public key and its user-specific private key. To improve efficiency, the client supports enabling parallel downloading and decryption, allowing multiple frames to be fetched while earlier frames are still being decrypted.
This feature is only applicable to the ABE workflow, since HTTPS already benefits from parallel cryptographic processing within the TLS pipeline.
By default, all downloading and decryption are performed entirely in memory, avoiding disk writes and eliminating unnecessary I/O overhead.

Using \stream, we compare three delivery modes: HTTP (no security), HTTP-ABE (selective coordinate decryption enabled), and HTTPS (Transport Layer Security).

In addition to streaming and decryption, \stream also performs detailed runtime logging. For each frame, the client records download time, decryption time, and the timestamped progression of playback events—including the current frame index, buffer occupancy, and buffer fill/empty transitions—to enable fine-grained analysis. \stream also detects and logs stalls, capturing both the start time and duration whenever playback pauses due to an empty buffer. These logs allow us to compute end-to-end streaming performance QoE metrics, such as rebuffering.

Note that \stream operates at the granularity of individual frames, representing a worst-case scenario where the number of GET requests is equal to the frame rate. Recent systems mitigate this cost using segment-based delivery, point cloud compression, and viewport adaptation as previously discussed in Sect.~\ref{subseec:relatedworks-streaming}. Our design isolates the effect of introducing an ABE layer on end-to-end performance, cacheability, and QoE. Optimizations in prior works are orthogonal to our contribution and can be incorporated in future work.


\section{Evaluation} \label{sec:evaluation}
In this section, we first describe our streaming setup and evaluation metrics, and then discuss the performance of \acro for point cloud video streaming in comparison to HTTPS and HTTP-only approaches.

\subsection{Streaming Setup \& Metrics} 
\label{subsec:vvsetup:metric}

\subsubsection{\textbf{Setup:}}

\paragraph{\textbf{Point cloud video dataset:}}
To emulate a 60-second point cloud video at 24 FPS, we generated $24 \times 60 = 1440$ point cloud frames using the same \textit{108k}-point cloud described earlier in Sect.~\ref{subsec:enc-impact}. This results in an approximate video bitrate of about 1~Gbit/s for our dataset. We then created four MPDs for our streaming experiments. One MPD was used for both HTTP-only and HTTPS streaming, listing the original, unencrypted point cloud frames. For HTTP-ABE, we prepared three additional MPDs—corresponding to the \texttt{XYZ}, \texttt{XY}, and \texttt{X} granularities—each referencing pre-encrypted frames generated using the respective selective-encryption scheme. We selected these three granularities because they provide meaningful visual obfuscation and significant computational benefits; beyond \texttt{X}, obfuscation becomes poor with no additional savings~\cite{10.1145/3704413.3765298}. This setup enables a direct comparison across HTTP-only, HTTP-ABE (three schemes), and HTTPS streaming.

\textbf{\textit{CloudLab setup:}} We perform our evaluations on the CloudLab testbed~\cite{duplyakin2019cloudlab} using a five-node topology. One node hosts the origin server containing the point cloud content, a second node acts as a cache directly connected to the server, and the remaining three nodes function as client machines, each connected directly to the cache. Each client node runs eight independent streaming clients, yielding a total of 24 clients.

Each cache–client link provides approximately 9.4~Gbit/s. Given our emulated point cloud video bitrate of 1~Gbit/s per stream, each client node requires up to 8~Gbit/s of bandwidth, meaning 9.4~Gbit/s per link is sufficient to avoid contention on client-side paths. In contrast, we deliberately allow the server–cache link to experience pressure: it also operates at 9.4~Gbit/s, serving all 24 clients whose combined demand is roughly 24~Gbit/s. This creates a congested server–cache link, consistent with prior streaming studies~\cite{10.1145/3712676.3714450,usmani2025securingimmersive360video}.

\textbf{\textit{Apache HTTP server and Apache Traffic server:}} We use the Apache HTTP Server~\cite{apache2} on the origin server with TLS/SSL enabled to support HTTPS. Caching is provided by Apache Traffic Server (ATS)~\cite{ATS}, which is configured using the remap plugin to forward all client requests to the origin. We evaluate two cache configurations: caching disabled (0~MB) and caching enabled with a 2000~MB cache size. For HTTPS experiments, ATS is configured to perform SSL termination, enabling secure connections between the origin and the cache and between the cache and each client.



\textbf{\textit{Streaming clients:}} On the client side, we run eight instances of the PC-Stream client on each of the three client nodes, totaling 24 clients. Each client uses a buffer size of 6, allowing six seconds of frames to be queued, which is also the initial startup buffer. For the HTTP-ABE experiments, we enable parallel download and decryption with a download queue size of 10. Client start times are scheduled using a Poisson distribution~\cite{poisson} with $\lambda = 5$, resulting in each client beginning its session approximately five seconds after the previous one, on average.


\subsubsection{\textbf{Evaluation Metrics:}}

\paragraph{\textbf{CPU load:}} We record CPU load on all nodes (origin server, cache, and client nodes) to compare performance across the HTTP-ABE granularities, HTTPS, and HTTP-only experiments. To do so, we use Linux’s \texttt{pidstat}~\cite{pidstat} to monitor CPU utilization at a one-second sampling interval and collect CPU usage over the full duration of each experiment.

    
\textbf{\textit{Hit Rate:}} We also compare the performance of the different approaches in terms of cache hit rates. We make use of ATS monitoring logs, which report the number of cache hits and misses during each experiment.



\textbf{\textit{Rebuffering:}} On the client side, we use rebuffering as a metric to compare QoE across the different approaches. Rebuffering represents the time during which playback stalls because the client’s buffer is empty. We obtain rebuffering data from the \stream client logs, which record stall duration for each frame. These per-frame values are accumulated to produce the total rebuffering time for each client session. We then average these totals across all clients and compute the rebuffering percentage relative to the full video duration (60 seconds) to assess overall performance.


\subsection{HTTP vs HTTP-ABE vs HTTPS}
\label{subsec:streaming}

The following evaluation assesses the performance of \acro using the end-to-end distribution pipeline using the setup and metrics defined in Sect.~\ref{subsec:vvsetup:metric}.

We analyze five streaming schemes: \textit{HTTP-ONLY} (no security), three HTTP-based \acro configurations—\textit{ABE-XYZ}, \textit{ABE-XY}, and \textit{ABE-X}—and \textit{HTTPS}, which secures communication with TLS.

For each scheme, we conduct three experiments: one with caching disabled (0~MB) and two with a 2000~MB cache. The first 2000~MB run serves as a warm-up to populate the cache, and no metrics are collected during this phase. The second run is performed right after that, and only its results are used for evaluation. This warm-up process removes cache start-up effects and ensures fair comparison across all schemes.





\subsubsection{\textbf{CPU Load:}}\label{subsubsec:cpuload} First, we discuss the average CPU load on the server, cache, and clients. Since each node on CloudLab has 32 logical threads (across 16 physical cores), we sum the usage on all and plot the average usage throughout each streaming session. This period covers the start and end of all streaming sessions initiated by the 24 clients. 

\begin{figure*}[]\centering
  \includegraphics[width=1\linewidth]{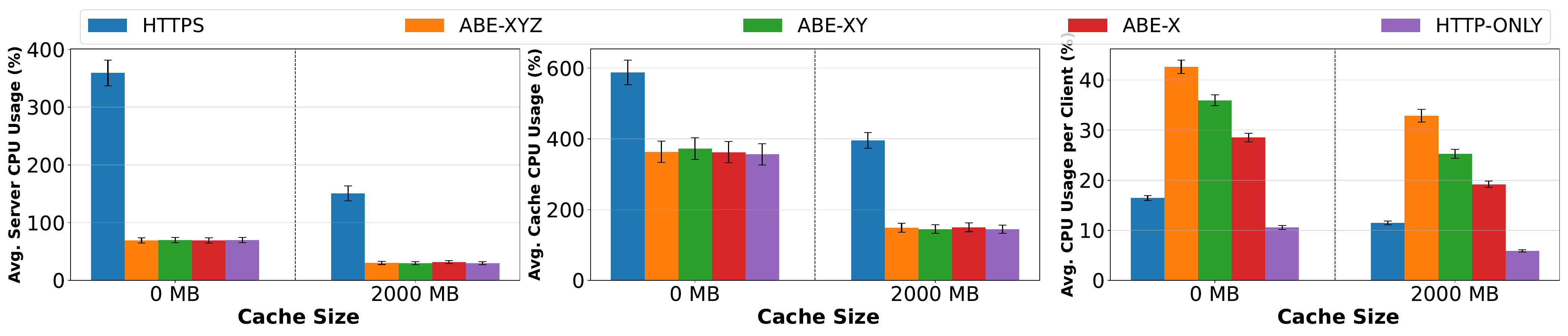}
  \captionsetup{justification=centering}
  \caption{Average CPU usage; server (left), cache (middle), and per client (right).}
    \label{fig:CPU:server-cache-client}
\end{figure*}

\textbf{Server CPU Load:} Figure~\ref{fig:CPU:server-cache-client} (left) shows the average CPU load at the origin server.
As expected, HTTPS incurs the highest CPU usage because the server must perform TLS encryption for every incoming request. In contrast, the ABE-based schemes operate on pre-encrypted segments—leveraging ABE’s ability to encrypt content once and distribute it securely to many users—thereby avoiding the repeated per-client encryption that TLS requires. Under the 0MB cache configuration, all HTTP-based ABE schemes—as well as the HTTP-only baseline—use approximately 80\% less CPU than HTTPS. With a 2000MB cache, the overall CPU load decreases across all schemes due to the reduction in forwarded requests from the cache. However, the relative gap remains consistent, with HTTPS still requiring significantly more CPU and the other schemes maintaining roughly an 80\% reduction in comparison.

\textbf{Cache CPU Load:} For CPU usage at the cache, we observe a trend similar to that of the origin server. Figure~\ref{fig:CPU:server-cache-client} (middle) shows the average CPU load on the cache for both the 0~MB and 2000~MB configurations. HTTPS consistently exhibits the highest CPU usage because the cache must perform TLS operations both when fetching content from the server and when serving it to clients. For the latter case, this is also true even if the content is already cached. As expected, overall CPU usage is higher in the 0~MB configuration for all schemes, since every request must be forwarded to the origin server. In this case, the HTTP-ABE and HTTP-only schemes use, on average, 36–39\% less CPU than HTTPS. With a 2000~MB cache, CPU usage decreases across all schemes due to fewer forwarded requests, but the relative difference widens: HTTP-ABE and HTTP-only reduce CPU consumption by roughly 63\% compared to HTTPS.

\textbf{Client CPU Load:}
Finally, we examine CPU usage on the clients. For each experiment, we average the total CPU utilization across all three client nodes and report per-client usage in Fig.~\ref{fig:CPU:server-cache-client} (right). Among all schemes, the ABE configurations exhibit the highest client-side CPU load—most notably ABE-XYZ. This is expected: client-side processing involves point cloud parsing, selective coordinate decryption, and reconstruction (Sect.~\ref{subsec:implement-enc}), and the workload grows with the number of coordinates targeted for decryption.

With the 0~MB cache configuration, ABE-XY reduces CPU usage by about 16\% compared to ABE-XYZ, while ABE-X reduces it by 33\%. HTTPS requires 61\% less CPU than ABE-XYZ, and HTTP-only uses 75\% less, since it involves no decryption. With a 2000~MB cache, overall CPU usage decreases for all schemes, and the relative differences shift slightly: ABE-XY uses 23\% less CPU than ABE-XYZ, ABE-X uses 41\% less, HTTPS saves 65\%, and HTTP-only saves 82\%. Across both cache sizes, HTTP-only consistently consumes the least CPU because no decryption is performed at the client.

\begin{figure*}[ht]\centering
  \includegraphics[width=1\linewidth]{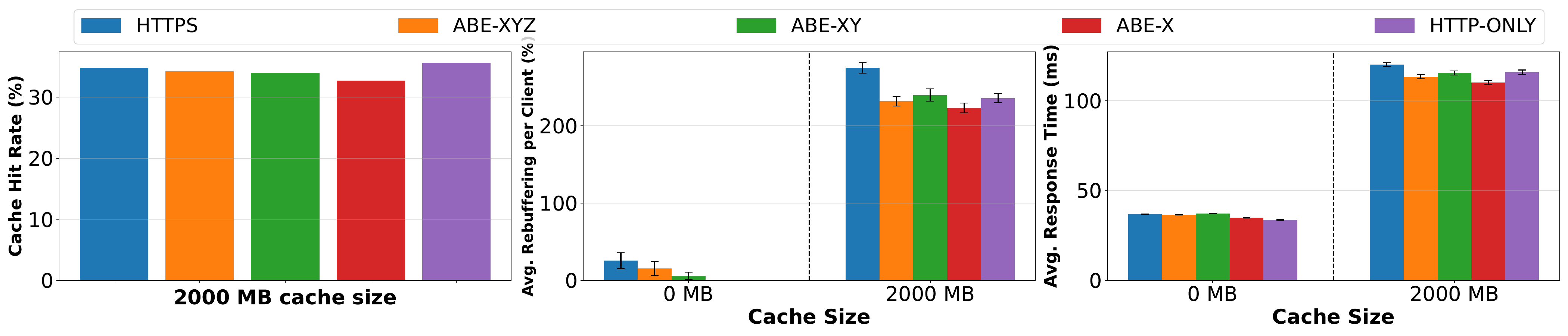}
  \captionsetup{justification=centering}
  \caption{Cache hitrates (left), average rebuffering per client (middle) and average cache response time per request (right).}
    \label{fig:hitrate-rebuffering-response}
\end{figure*}

\subsubsection{\textbf{Hitrates:}} \label{subsubsec:hits}
We also evaluate cache performance in terms of hit rates. Figure~\ref{fig:hitrate-rebuffering-response} (left) reports the hit rate percentages for the 2000~MB cache configuration across all schemes. The hit rate for the 0~MB/disable cache is always zero and is therefore omitted from analysis. All schemes achieve similar hit rates of roughly 33–35\%, indicating that secure streaming has little impact on cache effectiveness compared to HTTP-only. This includes the ABE schemes, even though the pre-encrypted frames are slightly larger in size due to ABE overhead.

\subsubsection{\textbf{Rebuffering:}} \label{subsubsec:rebuffering}
Figure~\ref{fig:hitrate-rebuffering-response} (middle) reports client-side QoE in terms of rebuffering. Rebuffering is expressed as a percentage of the total point cloud video duration (60 seconds).

For the 0~MB cache configuration, we observe the expected behavior: HTTPS experiences the highest rebuffering (about 25\%), followed by ABE-XYZ at roughly 15\% and ABE-XY at around 6\%. For ABE-X and HTTP-only, no rebuffering is observed. These results indicate that ABE-X is the lightest selective-encryption scheme—it provides reasonable obfuscation~\cite{10.1145/3704413.3765298} while adding minimal latency, resulting in no client-side QoE degradation. ABE-XY incurs only minor rebuffering, ABE-XYZ somewhat more, while HTTPS performs the worst, demonstrating that our ABE-based approach introduces substantially lower overhead than HTTPS.

Under the 2000~MB cache configuration, however, we observe an unexpected trend: rebuffering increases dramatically compared to the 0~MB case. While larger caches usually reduce stalls due to higher hit rates—as shown in prior studies~\cite{10.1007/978-3-319-32689-4_26,10.1145/3712676.3714450,usmani2025securingimmersive360video}—our experiments show rebuffering of 220–240\% for all ABE and HTTP-only schemes, and about 275\% for HTTPS. We also do not see the expected pattern where lighter ABE schemes (or HTTP-only) yield lower rebuffering. Upon investigation, we found that this anomaly is caused by increased cache response times under the 2000~MB configuration, which we discuss in the following subsection.

\subsubsection{\textbf{Cache Response Time:}}\label{subsubsec:response}
To understand the unexpectedly high rebuffering observed with the 2000~MB cache configuration, we examined the internal cache logs and analyzed per-request response times. Figure~\ref{fig:hitrate-rebuffering-response} (right) reports the average cache response time per request for all schemes under both the 0~MB and 2000~MB configurations. We found that cache response times with the 2000~MB cache were significantly higher than with the 0~MB configuration. Specifically, with 0~MB, average response times across all schemes ranged between 34–37 ms. In the case of a 2000~MB cache, they increased to 110–120 ms—roughly a threefold increase.

This behavior is due to the cache’s slower disk read/write operations. In the 0~MB configuration (or in cases of cache misses), the cache simply fetches content from the origin and forwards it directly from memory. In contrast, with the 2000~MB cache, the cache must first locate and then read each frame from disk before serving it to the client. These disk operations substantially increase latency, and the effect is amplified in our per-frame streaming setup, where at least 12 clients are simultaneously requesting point cloud frames at a rate of 24 FPS.

With a faster cache, such as a RAM-based cache, this bottleneck would be largely eliminated, and even with caching enabled at 2000~MB we would expect trends similar to those observed in the 0~MB configuration.
With our current 2000~MB case, the elevated cache response times dominate overall latency and lead to the large rebuffering values seen earlier. We also note that client CPU usage at 2000~MB is lower than in the 0~MB case (Fig.~\ref{fig:CPU:server-cache-client}, right). This is because higher latency and frequent stalls spread the same computational workload over a longer time window, thereby reducing the measured average CPU load.
To confirm our hypothesis, we repeat the experiments using a fast RAM-based cache in the subsequent section.

\subsection{RAM-based Caching: HTTP vs HTTP-ABE}

In this section we investigate whether the higher response times observed with caching enabled in prior experiments are caused by bottlenecks associated with disk-based caching. To this end, we repeat the experiments using \textbf{\textit{Varnish Cache}}~\cite{varnish}, a purely RAM-based caching system, placed between the origin server and clients, replacing the previously used \textbf{\textit{Apache Traffic Server (ATS)}}.

A limitation of Varnish is that it does not support HTTPS/TLS and operates exclusively over HTTP. Consequently, we restrict this evaluation to HTTP-only and HTTP-ABE configurations.
Nevertheless, HTTPS is known to impose additional overhead, and based on our earlier results with a disk-based ATS cache, we expect HTTPS to follow similar relative performance trends. Therefore, HTTP-based ABE approaches are likely to outperform HTTPS, and the conclusions drawn here remain applicable.

All results here are computed as the mean over five independent runs to provide a more robust estimate, and the error bars represent the 95\% confidence interval across these runs.


\begin{figure*}[]\centering
  \includegraphics[width=1\linewidth]{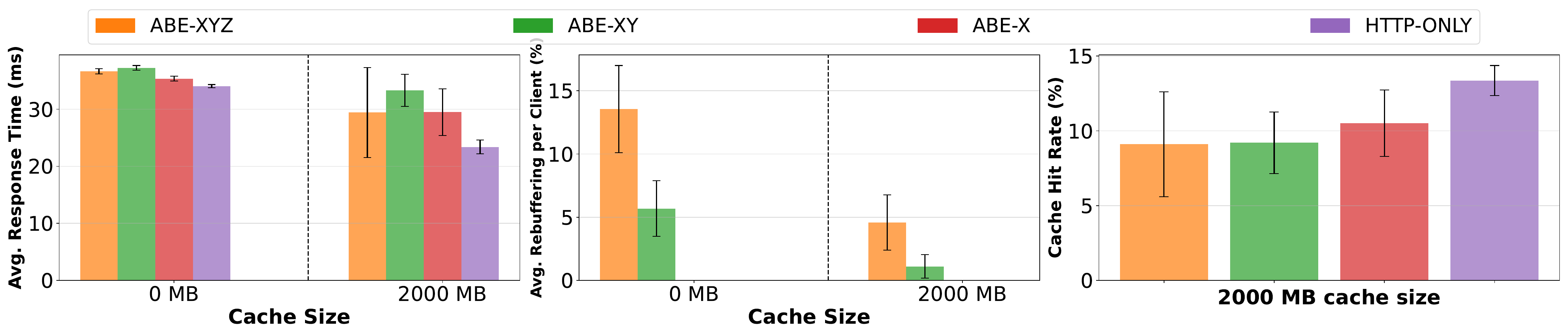}
  \captionsetup{justification=centering}
  \caption{RAM-based cache results: average response time(left), average rebuffering per client (middle), and hitrates (right).}
    \label{fig:Varnish-hitrate-rebuffering-response}
\end{figure*}

\begin{figure*}[]\centering
  \includegraphics[width=1\linewidth]{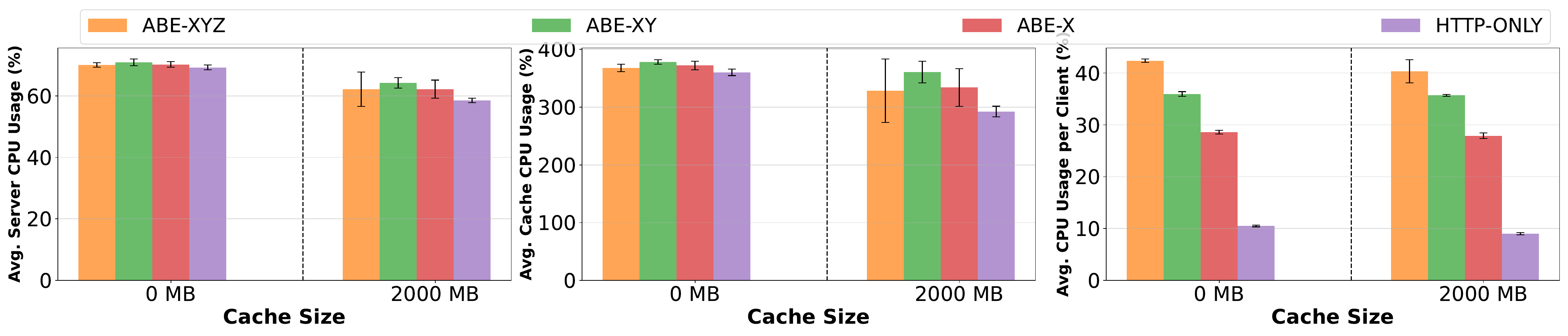}
  \captionsetup{justification=centering}
  \caption{RAM-based cache results: average CPU usage; server (left), cache (middle), and per client (right).}
    \label{fig:Varnish-CPU:server-cache-client}
\end{figure*}

\subsubsection{\textbf{Cache Response Time:}}
We first examine the average cache response time per request, shown in Fig.~\ref{fig:Varnish-hitrate-rebuffering-response} (left). In the 0~MB cache configuration, response times range between 34-37 ms across all schemes, consistent with observations from the disk-based ATS cache. In contrast, when caching is enabled with a 2000~MB configuration, average response times slightly decrease to approximately 29-33 ms for ABE-based schemes and to around 23 ms for HTTP-only traffic. This behavior differs from our ATS-based results, where enabling a 2000~MB cache increased response times. These findings confirm our earlier hypothesis that the elevated response times under ATS were primarily caused by disk read/write bottlenecks, which are eliminated by RAM-based Varnish cache.


\subsubsection{\textbf{Rebuffering:}}
Figure~\ref{fig:Varnish-hitrate-rebuffering-response} (middle) presents the average rebuffering ratios. For the 0~MB configuration, rebuffering remains around 13.5\% for ABE-XYZ and 4.5\% for ABE-XY, while no rebuffering is observed for ABE-X and HTTP-only schemes. These results closely mirror those obtained with ATS caching, consistent with the similar response times observed in the absence of caching.

When the 2000~MB cache is enabled, the reduced response times directly translate into lower rebuffering. With Varnish, we observe approximately 5.6\% rebuffering for ABE-XYZ, 1\% for ABE-XY, and no rebuffering for the remaining two schemes. This represents a substantial improvement over ATS-based caching, where rebuffering reached 220-240\% for both ABE and HTTP-only configurations.


\subsubsection{\textbf{Hitrates}}
Cache hit rates are shown in Fig.~\ref{fig:Varnish-hitrate-rebuffering-response} (right). With RAM-based Varnish caching, hit rates range between 9-13\% across schemes, which is notably lower than the 32-35\% observed with disk-based ATS. This reduction can be attributed to the lower response latency of Varnish. Faster cache responses reduce the temporal spacing between client requests, leaving fewer overlapping requests in flight for the same objects. As a result, the opportunity for cache reuse decreases, leading to lower observed hit rates. In contrast, the higher latency of disk-based ATS effectively serialized requests over longer intervals, increasing the likelihood of cache hits.

\subsubsection{\textbf{CPU Load:}}
\label{subsubsec:Varnish-cpuload}

Figure~\ref{fig:Varnish-CPU:server-cache-client} reports the average CPU utilization at the server, cache, and client. In the 0~MB configuration, CPU usage with RAM-based Varnish closely matches that observed with disk-based ATS across all components, consistent with the similar response times in this setting.

When caching is enabled with a 2000~MB configuration, we observe higher CPU utilization at the server, cache, and clients under Varnish for all schemes compared to ATS. At the server, CPU utilization reaches up to 64\% across schemes, compared to a maximum of 32\% with ATS. At the cache, utilization increases to 328-360\% for ABE-based schemes and 292\% for HTTP-only traffic, whereas ATS previously exhibited 145-150\% across all schemes. On the client side, CPU utilization remains quite similar to 0~MB configuration, specifically it reaches 28\% for ABE-X, 36\% for ABE-XY, and 40\% for ABE-XYZ, while HTTP-only traffic incurs only 9\%. In contrast, ATS resulted in 19-35\% utilization for ABE granularities and 6\% for HTTP-only traffic.

The higher average CPU utilization observed with RAM-based Varnish caching follows the same trend as the cache hit-rate behavior. Lower response times allow streaming downloads to complete within shorter time windows, reducing rebuffering and increasing delivery speed. While the total computational work remains unchanged, executing it over a shorter duration results in higher measured average CPU utilization. In contrast, ATS incurs higher response times and rebuffering, spreading the same workload over longer intervals and thus lowering average CPU usage. This interpretation is supported by the 0~MB configuration, where comparable response times between Varnish and ATS lead to similar CPU utilization.

\textit{Overall, the higher CPU utilization observed with RAM-based Varnish caching is a positive indicator of improved throughput, lower latency, and faster data processing rather than increased computational overhead.}

\section{Conclusions} \label{sec:conclusion}

In this work, we presented \acro, which enables selective coordinate encryption for protecting point cloud video streaming using Attribute-Based Encryption (ABE). Rather than encrypting entire frames, \acro encrypts only targeted subsets of the X, Y, and Z coordinates, which significantly lowers computational cost. Our results show that encrypting only the X coordinate offers up to 40\% lower encryption time and up to 75\% lower decryption time compared to full-frame encryption.



Our work is novel in integrating and evaluating security within the point cloud streaming pipeline. In our evaluations, we compared three ABE granularities (XYZ, XY, and X) against HTTPS and HTTP-only baselines on a CloudLab deployment. The results show that our ABE schemes significantly reduce CPU load at the server and cache by up to 80\% and 63\%, respectively, while maintaining cache hit rates comparable to unsecured HTTP-only streaming. On the client side, ABE-XYZ and ABE-XY incur lower rebuffering than HTTPS, while ABE-X achieves rebuffering-free playback similar to HTTP-only. Although our ABE schemes increase client-side CPU usage, the added overhead remains manageable. Finally, we show that adopting RAM-based caching mitigates the increased response time observed with disk-based caching, reducing rebuffering while preserving performance trends across all evaluated schemes.

\section{Future Works}
In this work, we evaluated \acro in an on-demand streaming setting. Since ABE supports point-to-multipoint distribution, all point cloud frames were pre-encrypted and stored at the origin server in encrypted form. As a result, no per-request encryption is required during streaming for the ABE schemes. In contrast, HTTPS relies on point-to-point TLS encryption, requiring the origin server to perform encryption separately for each incoming request.

For live streaming scenarios, ABE encryption would need to be performed at the origin server as new point cloud frames are generated. However, unlike HTTPS, ABE would still avoid repeated encryption of identical content for multiple concurrent viewers. As discussed in Sect.~\ref{subsec:enc-impact}, our evaluation of selective coordinate encryption shows that encrypting only the X coordinates reduces encryption time by up to 40\% compared to full-frame encryption. These results suggest that applying \acro to live volumetric streaming could offer meaningful performance benefits, making it an interesting direction for future work.

As discussed earlier in Sect.~\ref{subseec:relatedworks-streaming}, recent work has focused on optimizing volumetric video streaming through techniques such as point cloud compression, segment-based delivery, and adaptive bitrate (ABR) streaming. In contrast, our study focuses on per-frame delivery and single–point-cloud-level serving, representing a worst-case baseline for comparison. This opens several avenues for future work, including integrating point cloud compression and segment-level delivery, incorporating ABR mechanisms, and exploring selective encryption strategies at the compressed segment level for volumetric video streaming.

\balance
\bibliographystyle{ACM-Reference-Format}
\bibliography{refs, susmit}

\end{document}